\documentclass[twocolumn,prb]{revtex4}

\usepackage{graphicx}
\usepackage{dcolumn}
\usepackage{bm}
\usepackage{natbib}

\begin{document}

\preprint{APS/123-QED}

\title{Tunability and Losses of Mid-infrared Plasmonics in Heavily Doped Germanium Thin Films}

\author{Jacopo Frigerio, Andrea Ballabio, Giovanni Isella}
\affiliation{L-NESS, Dipartimento di Fisica, Politecnico di Milano, Polo di Como, Via Anzani 42, I-22100 Como, Italy}

\author{Emilie Sakat, Paolo Biagioni}
\affiliation{Dipartimento di Fisica, Politecnico di Milano, Piazza Leonardo da Vinci 30, I-20100 Milan, Italy}

\author{Monica Bollani}
\affiliation{CNR-IFN and L-NESS, Via Anzani 42, I-22100 Como, Italy}

\author{Enrico Napolitani}
\affiliation{Dipartimento di Fisica e Astronomia, Universit{\'a} di Padova and CNR-IMM MATIS, Via Marzolo 8, I-35131 Padova, Italy}

\author{Costanza Manganelli}
\affiliation{Scuola Superiore Sant’Anna, via G. Moruzzi 1, I-56124 Pisa, Italy}

\author{Michele Virgilio}
\affiliation{Dipartimento di Fisica "E. Fermi", Università di Pisa, Largo Pontecorvo 3, I-56127 Pisa, Italy}

\author{Alexander Grupp, Marco P. Fischer, Daniele Brida}
\affiliation{Department of Physics and Center for Applied Photonics, University of Konstanz, D-78457 Konstanz, Germany}

\author{Kevin Gallacher, Douglas J. Paul}
\affiliation{School of Engineering, University of Glasgow, Rankine Building, Oakfield Avenue, Glasgow G12 8LT, United Kingdom}

\author{Leonetta Baldassarre, Paolo Calvani, Valeria Giliberti, Alessandro Nucara and Michele Ortolani}
 \affiliation{Dipartimento di Fisica, Sapienza Universit{\'a} di Roma, Piazzale Aldo Moro 5, I-00185 Rome, Italy}
\email{michele.ortolani@roma1.infn.it}

\date{\today}

\begin{abstract}
Heavily-doped semiconductor films are very promising for application in mid-infrared plasmonic devices because the real part of their dielectric function is negative and broadly tunable in this wavelength range. In this work we investigate heavily n-type doped germanium epilayers grown on different substrates, in-situ doped in the $10^{17}$ to $10^{19}$ cm$^{-3}$ range, by infrared spectroscopy, first principle calculations, pump-probe spectroscopy and dc transport measurements to determine the relation between plasma edge and carrier density and to quantify mid-infrared plasmon losses. We demonstrate that the unscreened plasma frequency can be tuned in the 400 - 4800 cm$^{-1}$ range and that the average electron scattering rate,  dominated by scattering with optical phonons and charged impurities, increases almost linearly with frequency. We also found weak dependence of losses and tunability on the crystal defect density, on the inactivated dopant density and on the temperature down to 10 K. In films where the plasma was optically activated by pumping in the near-infrared, we found weak but significant dependence of relaxation times on the static doping level of the film. Our results suggest that plasmon decay times in the several-picosecond range can be obtained in n-type germanium thin films grown on silicon substrates hence allowing for underdamped mid-infrared plasma oscillations at room temperature. 
\end{abstract}

\pacs{Valid PACS appear here}
\maketitle

The recent push towards applications of spectroscopy for chemical and biological sensing in the mid-infrared (mid-IR) \citep{soref2010, faist2007, neubrech2008, adato2009, olmon2010, giannini2011, dandrea2013, cubukcu2013} has prompted the need for conducting thin films displaying values of the complex dielectric function $\tilde{\epsilon}(\omega) = \epsilon^{\prime}(\omega)+i\epsilon^{\prime\prime}(\omega)$ that can be tailored to meet the needs of novel electromagnetic designs exploiting the concepts of metamaterials, transformation optics and plasmonics \citep{maier2001}. In the design of metamaterials, where subwavelength sized conducting elements are embedded in dielectric matrices, if the values of $\epsilon^{\prime}$ of the metal and the dielectric are of the same order, but have opposite sign, the geometric filling fractions of the metal and dielectric can be readily tuned to achieve subwavelength-resolution focusing of radiation \citep{pendry2000}. Such requirement is met by silver for wavelengths $\lambda$ around 400 nm. The same condition cannot be achieved in the IR range by using elemental metals, however, because metals possess an extremely high negative value of $\epsilon^{\prime}$ not equaled, in absolute value on the positive side, by any dielectric material. It has been proposed that, in order to obtain tunable values of $\epsilon^{\prime}$ in the entire IR range \citep{boltasseva2011, BoltassevaRev}, heavily doped semiconductors \citep{hoffman2007, ginn2011, shahzad2011, li2011, law2012, law2013, baldassarre2015} and conducting oxides \citep{OxideRev} may be used  because their free carrier density can be set by selecting the doping level and further tuned by electrostatic gating \citep{li2011} or optical excitation \citep{berrier2010, wagner2014}. 

\begin{figure}
\includegraphics{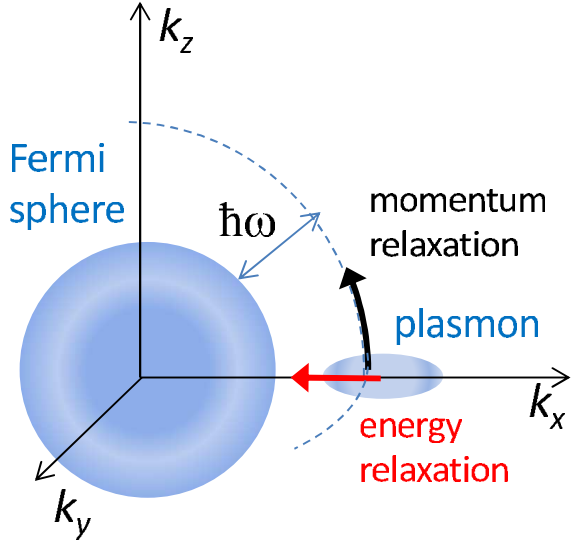}
\caption{\label{fig:plasmon} A schematic view in the k-space of a classical plasmon of frequency $\omega$ propagating along the $x$ direction, in the case of an isotropic three-dimensional electron liquid. The blue areas represent occupied electron states. Two types of decay channels exist for the plasmon: momentum relaxation by elastic scattering of electrons (black arrow) and energy relaxation by inelastic scattering (red arrow). Many electron scattering events of the two types may be needed for complete plasmon decay.}
\end{figure}

Beyond dielectric function tunability, a key requirement in plasmonics and metamaterial design is a low level of intrinsic losses in the material. Specific effects such as nanofocusing, field confinement or phase front shaping are obtained by creating subwavelength geometrical structures to engineer the propagation of a specific mode of the electromagnetic field. Any physical process, which leads to modification of the electron energy and/or the momentum distribution corresponding to the specifically engineered electromagnetic field mode, contributes to plasmon losses (see scheme in Fig.\ \ref{fig:plasmon}). For a given material, it is possible to make an assessment on the average single-electron scattering rate, which is an intrinsic property, but not on the plasmon decay time, which may be also dependent on the specific geometry, the dielectric environment, the surface roughness and/or the finite size of the subwavelength elements. At IR frequencies, intrinsic losses are due to non-radiative decay of interband transitions and intraband free-carrier excitations. Excluding perhaps superconductors, which can be used only in the microwave and terahertz ranges \citep{Anlage2011, Anlage2013, Limaj2014}, low intrinsic losses are hard to achieve in all classes of materials that have been considered for plasmonics and metamaterial applications. Interband transitions can be avoided only in material-specific frequency ranges; metal films deposited by standard techniques such as evaporation or sputtering display polycrystalline structures that increase finite size and surface roughness effects; conducting oxides are characterized by high crystal defect densities \citep{OxideRev}; compound semiconductors have strong dipole-active optical phonons that both directly absorb IR radiation and efficiently scatter the conduction electrons; finally, doped materials in general present charged-impurity densities that increase proportionally to the doping levels and produce Coulomb scattering of free carriers. Energy transfer from the engineered electromagnetic field mode to electrons and polar phonons in the material is quantified by $\epsilon^{\prime\prime}(\omega)>0$, which can therefore be taken as a figure of merit for intrinsic losses. It is worth noting that the only materials showing a wide range of negative values of $\epsilon^{\prime}(\omega)$ together with almost vanishing $\epsilon^{\prime\prime}(\omega)$ are surface-phonon-polariton materials, like silicon carbide \citep{Taubner} and boron nitride \citep{Maier}, where the negative value of $\epsilon^{\prime}(\omega)$ is not due to free carriers but to IR-active phonons. These materials, however, can operate in a very narrow and non-tunable range between the end of the transverse optical phonon absorption tail and the longitudinal optical phonon frequency, which is ultimately defined by the crystal structure and certainly non-tunable.

A further requirement of materials for applications in IR plasmonics is the technological potential: (i) it should be possible to grow or deposit them with a cost-effective and reliable technique; (ii) their processing should be compatible with mainstream microfabrication technology, in particular silicon foundry processes. The ensemble of these considerations has led us to study the electrodynamic properties of heavily-doped Ge thin-films grown by Chemical Vapor Deposition (CVD) on different substrates, including both ideal lattice-matched bulk Ge wafers and lattice-mismatched silicon wafers of high technological value. The present material platform is intended for designing metamaterials and plasmonic devices in the mid-IR range, here defined as the wavelength range $20 > \lambda > 5$ $\mu$m, or the electromagnetic frequency range $500 < \omega < 2000$ cm$^{-1}$, or the photon energy range $60 < \hbar\omega < 250$ meV. The high-frequency limit for conducting behavior of a given material with free carrier density $n$ and effective mass $m^*$ is ultimately set by the \emph{screened plasma frequency} or the frequency $\omega^*$ such that $\epsilon^\prime(\omega^*)=0$, being negative for lower $\omega$, approximately given (in the limit of low losses) by:

\begin{equation}
\label{screened}
\omega^* \simeq \sqrt{\frac{4\pi n e^2}{\epsilon_\infty m^*}}.
\end{equation}

\noindent where $e$ is the free carrier charge and $\epsilon_\infty$ is the high-frequency dielectric screening constant, which can be thought of as the square of the refractive index $\eta$ of the corresponding undoped material. In a three-dimensional conductor, $\omega^*$ approximately scales with the free carrier density $n$ as $n^{0.5}$, which then becomes the main optimization parameter for electromagnetic design. This work is devoted to a study of IR plasmonics in n-doped germanium (n-Ge) thin films, which display ideal intrinsic properties including $n$ spanning two orders of magnitude, small $m^*$ and $\omega^*$ reaching values well into the mid-IR. We present the experimental determination of the IR dielectric function of Ge thin films grown by CVD on different substrates. We demonstrate that this material displays both wide tunability of $\epsilon^\prime(\omega)$ and values of $\epsilon^{\prime\prime}(\omega)$, limited only by the fundamental quantum processes of electron-phonon and electron-charged impurity scattering. In order to quantify the role of the different loss mechanisms, we have developed a quantum model of electron scattering rates as a function of the plasmon frequency and we have conducted time-resolved pump-probe experiments to estimate its characteristic energy relaxation time. Although both the experiments and the modeling are specific to germanium, the final conclusions we draw on plasmonic losses can be extended to other semiconductors and doped materials in general.

\begin{figure}
\includegraphics{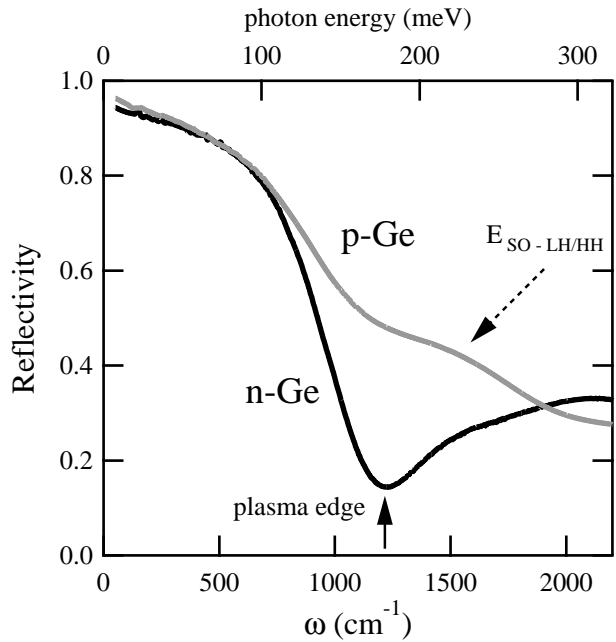}
\caption{\label{fig:pn}reflectivity of two Ge films grown on Si wafers. One film was hole-doped (p-Ge) and one electron-doped (n-Ge). The thickness is 1 $\mu$m and the free carrier density is $2\cdot10^{19}$cm$^{-3}$ for electrons and $7\cdot10^{19}$cm$^{-3}$ for holes.}
\end{figure}

Germanium is a close-to-ideal material for mid-IR optics. The main impurity-state transitions lie in the far-IR range, therefore lightly doped films display almost no absorption in the entire mid-IR range and a frequency-independent value of $\eta=4.0$ (then $\epsilon_\infty = 16.0$). The fundamental energy gap is indirect and equal to $E_g = 0.66$ eV, while the first direct interband transition edge is at $E_0=0.89$ eV at low temperatures, and this leads to an increase of the optical absorption above $\sim 6000 $ cm$^{-1}$, i.e. well above the mid-IR range of interest for molecular sensing. These features result in the abrupt decrease of the normal-incidence mid-IR reflectivity of n-Ge films from a value close to 1 at low frequencies to an absolute minimum just above $\omega^*$ (\emph{plasma edge}, see Fig.\ \ref{fig:pn}) where $\epsilon^{\prime}\sim 1$ and refractive-index matching with vacuum suppresses the reflected intensity. The conduction band minimum of Ge is at the $L$-point of the electron wavevector $\bf{k}$-space (intercepts of the $<111>$ axes with the first Brillouin zone) which leads to a Fermi surface (for n-doped materials) made of four ellipsoids with two short axes (corresponding to the small transverse effective mass $m_T = 0.0815 m_e$ ) and one long axis (large longitudinal effective mass $m_L = 1.59 m_e $). In the limit of vanishing momentum acquired by the electrons from the electric field, which is of interest for the isotropic optical and dc conductivities, the relevant quantity is the so-called \emph{conductivity effective mass} $m^*=0.12 m_e$ calculated as described in Ref.\ \onlinecite{Sze}. Since $\omega^*$ scales with $1/\sqrt{m^*}$, the relatively low value of $m^*$ makes n-Ge appealing for mid-IR plasmonic applications if compared to large $m^*$ materials like electron-doped Si ($m^*=0.26m_e$), Al-doped ZnO ($m^*=0.29m_e$), or In$_{1-x}$Ti$_x$O ($m^*=0.35m_e$), because, for a given value of the free carrier density $n$ (hence of the losses), a higher $\omega^*$ can be obtained. 

One may recall that bands with small $m^*$ can be found in hole-doped Ge (p-Ge) if compared to n-Ge \citep{levinger1961}. The presence of multiple electronic bands in the energy range of the valence band of cubic semiconductors with the diamond-like structure (including Ge), however, implies the existence of interband transitions (leading to plasmon losses) between the different valence bands in the entire IR range, not fully inhibited by the interband dipole selection rule. In particular, the transition between the light-hole (LH) and the heavy-hole (HH) band around the $\Gamma$ point take place in a broad frequency range in the far-IR ($10 < E_{LH-HH} < 100$ meV) due to the different effective masses of the two hole types that makes the two bands non-parallel in $\bf{k}$-space. The transitions from the split-off (SO) band to both the LH and HH bands is also activated by hole doping, is very broad and sits at $E_{\text{SO-LH/HH}} \sim 289$ meV \citep{Madelung}, exactly in the mid-IR range of technological interest. Instead, for n-Ge the only inter-conduction band transition is the diagonal L-to-$\Gamma$ transition at $E \sim 140$ meV, which requires a high momentum exchange with the lattice and has negligible optical spectral weight. These predictions are confirmed by the IR reflectivity of p-Ge thin films epitaxially grown on silicon wafers. In Fig.\ \ref{fig:pn}, the absolute reflectivity spectrum of p-Ge demonstrates a strong signature of the SO-LH/HH transitions between 150 and 250 meV (indicated by dotted arrow), to be contrasted with the plasma edge clearly observed in the reflectivity spectrum of n-Ge, which is indicative of a pure free-carrier response with neither interband nor impurity state transitions in the mid-IR. To summarize, inter-valence band transitions in the entire IR range produce doping- and $\omega$-dependent contributions to $\epsilon^{\prime\prime}(\omega)$ that would eventually undermine any design attempt for plasmonics in p-Ge. Dramatically, the strength of such lossy interband transitions increases with increasing hole-doping, because more final empty states become available. Instead, n-Ge displays all the \emph{a priori} characteristics to perform as an ideal mid-IR plasmonic material. It remains to be determined how broadly the dielectric function of n-Ge can be tuned and how much the unavoidable free-carrier losses in doped semiconductors will impact on the mid-IR plasmonic performance. This is the subject of the present paper.

The paper is organized as follows. In the first section we describe the n-Ge thin-film growth and the tunability range of their plasma frequency by selection of the doping level. In the second section we analyze the absolute reflectivity spectra in the entire far-to-near IR and we discuss the limitations of the commonly employed Drude-Lorentz model of the dielectric function based on a frequency-independent electron scattering rate. As an alternative method of loss evaluation, we calculate the dielectric function and the frequency-dependent electron scattering rate in a model-independent way exploiting the Kramers-Kronig relations. In the third section, we perform first-principle calculations of the electron scattering as a function of temperature, doping and frequency, including the effect of phonons and charged impurities, which can be used for evaluation of the losses in the mid-IR plasmonics. In the fourth section we corroborate the steady-state spectroscopy analysis and the first-principle calculations by a direct measurement of the collective energy-relaxation time in optically-pumped n-Ge films. Finally, we draw some conclusions on the use of doped semiconductors for future mid-IR plasmonic applications.

\begin{table*}
\caption{\label{tab:table1}Carrier density parameters}
\begin{ruledtabular}
\begin{tabular}{cccccccccc}
 Sample&PH$_3$ flux&substrate
&d($\mu$m)& $d_{\text{IR}}$($\mu$m) & $[P]$ (cm$^{-3}$) & $\rho_0 (\Omega$cm) & $n_{H}$ (cm$^{-3}$) & $n_{\text{IR}}$ (cm$^{-3}$) & $n_{\text{IR}}$/[P] \\ \hline
 8648&$1\%$&$\text{Si wafer}$&$1.39$ & $1.30$ & $2.0\cdot 10^{17}$ & $4.5\cdot10^{-2}$ & $1.4\cdot 10^{17}$ & $2.1\cdot 10^{17}$& 100\%\\
 8643&$10\%$&$\text{Si wafer}$&$1.45$ & $1.33$ & $3.6\cdot 10^{18}$ & $2.5\cdot10^{-3}$ & $3.4\cdot 10^{18}$ & $3.5\cdot 10^{18}$& 97\% \\
 8649&$25\%$&$\text{Si wafer}$&$1.18$& $1.00$ & $1.3\cdot 10^{19}$ & $1.4\cdot10^{-3}$ & $1.0\cdot 10^{19}$ & $1.1\cdot 10^{19}$& 85\% \\
 8644&$30\%$&$\text{Si wafer}$&$1.15$& $0.97$ & $2.1\cdot 10^{19}$ & $1.4\cdot10^{-3}$ & $1.5\cdot 10^{19}$ & $1.5\cdot 10^{19}$& 71\% \\ 
 9007&$50\%$&$\text{Si wafer}$&$1.00$& $0.99$ & $7.5\cdot 10^{19}$ & $5.5\cdot10^{-4}$ & $2.6\cdot 10^{19}$ & $2.5\cdot 10^{19}$& 33\% \\ \hline
 9332&$25\%$&$\text{Si wafer}$&$2.0$& $2.5$ & $1.3\cdot 10^{19}$ & $8.1\cdot10^{-4}$ & $0.9\cdot 10^{19}$ & $0.8\cdot 10^{19}$& 62\%  \\
 9338&$40\%$&$\text{Si wafer}$&$2.0$& $2.3$ & $3.5\cdot 10^{19}$& $4.3\cdot10^{-4}$ & $2.5\cdot 10^{19}$ & $2.3\cdot 10^{19}$& 66\%  \\
 9335&$40\%$&$\text{virtual Ge}$&$2.0$& $2.1$ & $3.5\cdot 10^{19}$ & $5.9\cdot10^{-4}$ & $2.9\cdot 10^{19}$ & $2.5\cdot 10^{19}$& 71\% \\
 9336&$40\%$&$\text{Ge wafer}$&$2.0$& $2.1$ & $3.5\cdot 10^{19}$ & $9.1\cdot10^{-4}$ & $3.3\cdot 10^{19}$ & $3.0\cdot 10^{19}$& 86\% \\
\end{tabular}
\end{ruledtabular}
\end{table*}

\section{Growth of Heavily Doped Films}

Different dopants can be incorporated into the lattice sites of Ge, but since we employ silicon-foundry compatible CVD techniques requiring non metal-organic gas precursors the choice is limited to B for hole doping and P and As for electron doping. In this work we will focus on films doped by phosphorous atoms since this is expected to give higher electron concentrations \citep{vanhellemont2012}. Indeed Ge epilayers doped with P have displayed carrier densities exceeding $10^{20}$ cm$^{-3}$. \citep{scappucci2015}. From here on, the word \emph{doping} indicates the free electron concentration $n$. We will analyze the effect of the incorporation opf P atoms in the Ge lattice on $n$ and on the electron scattering rate, leaving the discussion of the modification of structural, mechanical and optical properties to specific studies. 
We refer to donor atoms which effectively contribute to an increase in the charge-carrier density in the material as \emph{activated dopants} (i.e. ionized) with volume density $N_A$, while  \emph{inactivated dopants} with density $N_I$ designate P atoms that are incorporated into the crystal structure but do not contribute any free carrier. Secondary Ion Mass Spectrometry (SIMS) is used to determine the total P atom concentration $\left[P\right] = N_A + N_I$ but transport or optical techniques are required to measure the activation ratio $N_A/[P]$. Basic theory of solid-state solutions tells us that active dopants are those that substitute to Ge in a lattice position, while donor atoms that occupy interstitial sites or phase-separate into clusters are usually inactivated. It is well known that at high doping levels other effects take place, among them dopant-dimer formation \citep{scappucci2015}, increase of the dislocation density and clustering of dopants around crystal defects \citep{geiger2014}. All these effects contribute to the decrease of $N_A/[P]$. Inactivated dopants generally act as neutral impurities weakly contributing to scattering of free carriers. On the other hand, in principle $N_A = n$ and, since for mid-IR plasmonics $n$ requires to be as high as possible, these unavoidable charged impurities represent a major contribution to electron scattering. 

The thin-film growth  is performed in a low-energy plasma-enhanced chemical vapor deposition (LEPE-CVD) reactor \citep{rosenblad1998}. Therein, argon gas is introduced into the growth chamber after passing through a plasma source, where a tantalum filament is heated for thermionic emission. A dc arc discharge of 30 to 50 A is sustained between the filament and the growth chamber with a low voltage of 30 V and stabilized by an anode ring mounted in the growth chamber. Magnetic fields are used to focus the plasma onto the substrate heated up to a temperature $T_{\text{sub}}$. The deposition chamber has a base pressure of 10$^{-9}$ mbar, while the working pressure reaches 10$^{-2}$  mbar. The flow of precursors gases (GeH$_4$ for germanium, PH$_3$ and B$_2$H$_6$ for dopants, diluted in Ar) introduced in the chamber is regulated by mass flow controllers. The growth rate, controlled by the plasma density and by the flow of process gas, and the mobility of the adatoms, controlled by $T_{\text{sub}}$, can be optimized separately.  A list of the samples grown and characterized in this work is reported in \ref{tab:table1}. In this work we used a growth mode featuring arc discharge current of 30 A, magnetic field of 1 mT, $T_{\text{sub}} = 500 ^o$C and GeH$_4$ flux kept at 20 sccm. In the 864x sample series, grown on $\left[100\right]$ silicon wafers, the PH$_3$ flux was varied in percentage of its maximum value around 1 sccm. In the 933x series, the PH$_3$ flux was kept at 25\% or 40\% and three types of substrates were employed: (i) [100] germanium wafer for homo-epitaxial growth resulting in threading dislocation densities in the n-Ge films $< 10^{5}$cm$^{-2}$; (ii) [100] silicon wafers for hetero-epitaxial growth (due to the 4.2\% difference between the lattice constants of Si and Ge) of fully relaxed n-Ge films with high dislocation densities of the order of $10^{9}$cm$^{-2}$; (iii) the same [100] silicon wafers with a 2 $\mu$m thick undoped Ge layer (so-called \emph{virtual substrate}) which is cyclically annealed between 600 and 780 $^o$C to reduce the dislocation density down to $10^{7}$cm$^{-2}$ before growing the n-doped film \citep{osmond2009}. Indeed, heavily doped n-Ge films cannot be annealed at these temperatures because of the tendency of P atoms to form clusters \citep{geiger2014}. SIMS measurements have been performed to quantify the total P-atom incorporation by using a CAMECA ims-4f mass spectrometer. An O$_2^{+}$ ion beam with accelerating voltage of 12.5 keV, rastered over a $250\times250$ $\mu$m$^2$ area, was used for sputtering, while collecting $^{31}$P$^{16}$O$^+$ secondary ions. Calibration of $[P]$ was performed by measuring a Ge standard with known P areal density with an accuracy of $\pm$15\%. The concentration levels within the n-Ge films resulted to be uniform to within $\pm$5\%. 

\begin{figure}
\includegraphics{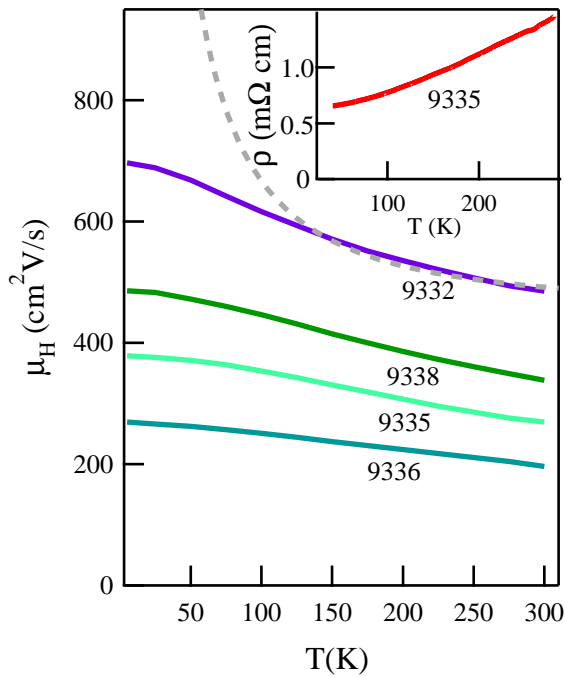}
\caption{\label{fig:hall} The Hall mobility of n-Ge films of thickness $\sim 2$ $\mu$m and varying substrate type and doping level (see Table \ref{tab:table1}). The Hall mobility decreases with doping indicating the key role of electron scattering from charged impurities. The grey dashed line represents the theoretical $T^{-\frac{3}{2}}$ behavior for sample 9332. In the inset, the resistivity vs. temperature curve of sample 9335 is reported.}
\end{figure}

Temperature ($T$)-dependent dc transport measurements have been performed  in order to determine the dc resistivity $\rho(T)$ and the Hall coefficient $R_H(T)$. As shallow defects can be thermally activated even at room $T$ to increase the measured transport carrier density, it is important to undertake $T$-dependent Hall measurements to determine the activated carrier density. For samples from the 933x series, data were acquired at all $T$ from 5 to 300 K with a step of 0.1 K/s. For the 864x series, the measurements were performed only at $T=300$ K. All dc transport data were collected from six-terminal Hall bars processed by UV lithography and dry etching. A Ni/Ti/Al (100 nm/5 nm/100 nm) metal stack was evaporated on the electrical pads and then annealed at 340 $^o$C for 30 s in order to create low-resistivity ohmic contacts \citep{gallacher2012}. The temperature dependence of $\rho(T)$ has been found to be almost linearly dependent on $T$ from 300 K to 50 K for all samples (see inset of Fig.\ \ref{fig:hall}). $R_H(T)$ measured applying a magnetic field of $2.5$ T displayed a constant value, different for each sample, over the full temperature range. This is expected, because all samples of the 933x series are doped close or beyond the level of the Mott transition in Ge ($N_{Mott} \simeq 2.5 \cdot 10^{17}$cm$^{-3}$) \citep{Mott1, Mott2, Mott3}. The measured value of $R_H$ was used to determine the Hall carrier density $n_{H}=-1/eR_H$ reported in Table \ref{tab:table1}. The Hall mobility $\mu_H(T) = (R_H\rho)^{-1}$ calculated at all $T$ is shown in Fig.\ \ref{fig:hall}. $\mu_H$ mostly depends on the electron mobility, because at the high electron-doping levels considered here the contribution of holes to dc transport is negligible. The evaluation of the electron mobility $\mu$ from the Hall mobility $\mu_H$ requires in principle an accurate knowledge of the geometrical Hall factor \citep{hallfactor}, however the value of $\mu_H$ as a function of $T$ and $n$ can be used for a first-step evaluation of losses in mid-IR plasmonics, as is often undertaken in the literature \citep{BoltassevaRev}.

In lightly-doped group-IV elemental semiconductors (Ge and Si), the temperature dependence of the Hall mobility in the high-$T$ range above 100 K is dominated by the deformation potential scattering of free carriers because of the lack of polar optical phonons \citep{PaulRev}. The power law expected by taking into account the phonon population (linearly increasing with $T$) and the average thermal velocity of electrons (increasing as $T^{\frac{1}{2}}$)  is $\mu_H(T)\sim T^{-\frac{3}{2}}$ \citep{Mirza2014}, but the behavior of our samples significantly deviates from this law, especially at low $T$ (see sample 9332 in Fig.\ \ref{fig:hall}). The reason for this deviation is to be found in the key role of charged impurity scattering, also confirmed by the high value of the residual zero-temperature resistivity $\rho_0$ which increases with $n$ (see Table \ref{tab:table1}). The Hall scattering time can be calculated as $\tau_H=\mu_H m^* / e$ which is usually interpreted as the average time between subsequent electron scattering events and typically falls in the sub-picosecond range. A more rigorous interpretation of the dc transport data of Fig.\ \ref{fig:hall} in terms of fundamental electron scattering mechanisms is not possible at this stage, because the contribution of the non-insulating substrates cannot be precisely subtracted especially in the high temperature range $T \agt 100$ K \citep{hallfactor}. In any case, a thorough discussion of electron scattering and losses in n-Ge plasmonic resonators cannot be based on dc transport only, and a direct determination of the dielectric function at IR frequencies is needed.

\begin{figure}
\includegraphics{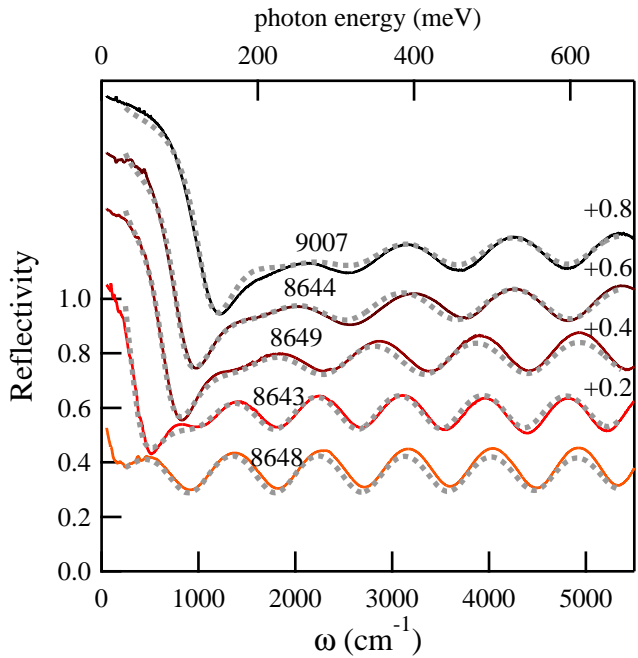}
\caption{\label{fig:863x}The absolute reflectivity at room $T$ of n-Ge thin films of thickness $\sim 1$ $\mu$m, grown on silicon wafers with varying PH$_3$ flux (see Table \ref{tab:table1}). The dashed lines are the best-fit to the Drude-Lorentz model for the dielectric function of n-Ge (Eq.\ \ref{model}). Curves are offset for clarity, the offset value is indicated on the right.}
\end{figure}

\begin{figure}
\includegraphics{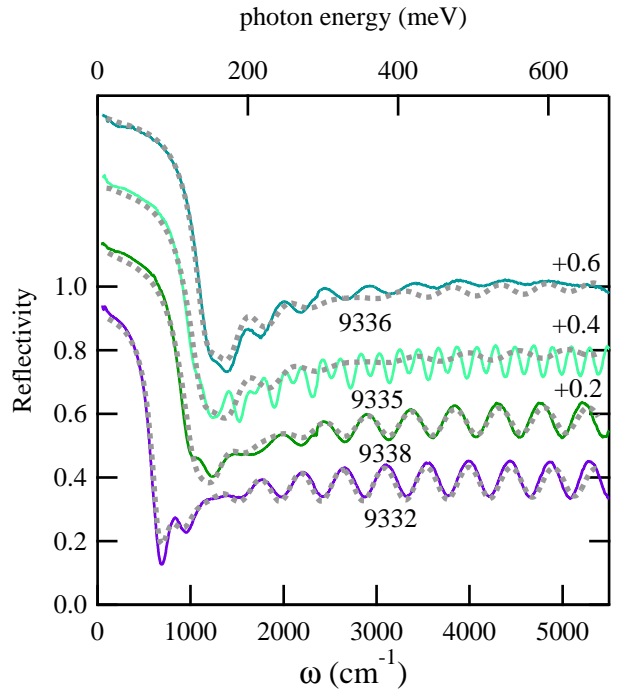}
\caption{\label{fig:933x}The absolute reflectivity at room $T$ of n-Ge thin films of thickness $\sim 2$ $\mu$m grown on different substrates (see Table \ref{tab:table1}). The dashed lines are the best-fit to the Drude-Lorentz model for the dielectric function of n-Ge (Eq.\ \ref{model}). The  model does not include the virtual substrate interface to the wafer, therefore it cannot reproduce the oscillations seen in the data of sample 9335 above $\sim 1500$ cm$^{-1}$, but this is inessential for the determination of the free carrier parameters. Curves are offset for clarity, the offset value is indicated on the right.}
\end{figure}

\section{Dielectric Function Determination}

\begin{figure}
\includegraphics{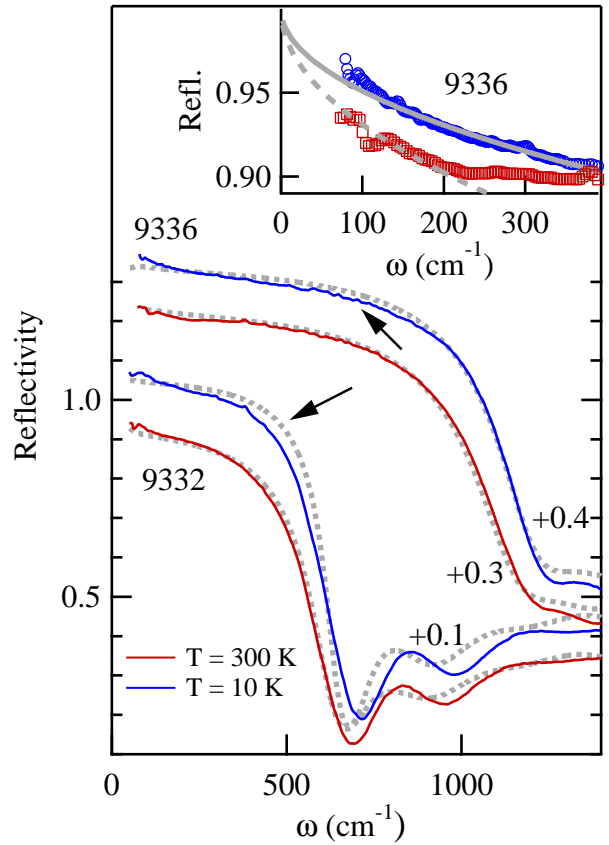}
\caption{\label{fig:drude}(Color online). Drude-Lorentz fitting (dotted grey lines) is possible for the room-$T$ reflectivity data (red, lower offset) but not possible for the low-$T$ data (blue, higher offset): the black arrows indicate the frequency ranges where the data and the model do not overlap. The offset values are indicated on the right. In the inset, enlarged view of the far-IR data of sample 9336 at $T=10$ K (circles) and at $T=300$ K (squares) with the Hagen-Rubens fit (thick grey lines).}
\end{figure}

Reflection spectroscopy performed at all $\omega$ from the far-IR to the near-IR provides a wealth of information on heavily doped semiconductor thin films, which complements the transport properties. The $\omega$- and $T$-dependent absolute normal-incidence reflectivity $R(\omega)$ was measured in the $50<\omega< 6000$ cm$^{-1}$ range with a Michelson-type Fourier-Transform spectrometer (Bruker IFS66v) equipped with a suite of beamsplitters and cryogenic detectors: a HgCdTe photovoltaic detector from Infrared Associates, whose response was linearized for different power levels, and a 4 K silicon bolometer from Infrared Labs, which features linear responsivity. The beam from the Michelson interferometer is sent to a homemade purely normal-incidence reflectivity setup based on the insertion of a broadband beamsplitter that conveys to the detectors around half of the beam reflected by the sample, which was glued on a copper ring and kept in a liquid-helium flow cryostat. Reference spectra were acquired at all $T$ and $\omega$ on a gold mirror, also kept in the cryostat and displaced exactly at the position previously occupied by the sample. The absolute reflectivity of all samples and also of bare substrates was obtained by dividing the sample spectrum by the reference spectrum taken with the same optical path alignment. The merged $R(\omega)$ datasets for the 864x and 933x series at room $T$ are shown in Fig.\ \ref{fig:863x} and Fig.\ \ref{fig:933x}, respectively. Besides the sinusoidal oscillations due to Fabry-Perot interference in the Ge film, one sees that the plasma edge moves to higher frequency for higher $n$ as expected from Eq.\ \ref{screened}. In Fig.\ \ref{fig:drude} the low-$T$ spectrum is also shown for two samples: the position of the plasma edge does not change appreciably with cooling, in agreement with the measured $T$-independent value of $n_H(T)$.

The silicon wafers in use in the electronic industry, here employed as substrates for CVD growth, display different types of IR-active impurity state transitions that severely impact on the transmission spectrum of the multilayer formed by the film and the much thicker substrate. For this reason, in our experiment we measure instead $R(\omega)$ which, for thick enough films is mainly determined by the optical properties of the CVD-grown n-Ge film. From the Fresnel relations at exactly normal incidence, we can write:

\begin{equation} 
\label{model}
R(\omega) \simeq \left|\frac{r_{12}+r_{23}e^{4\pi i \eta(\omega)d_{\text{IR}}}e^{-\alpha(\omega)d_{\text{IR}}}}{1 + r_{12}r_{23}e^{4\pi i \eta(\omega)d_{\text{IR}}}e^{-\alpha(\omega)d_{\text{IR}}}}\right|^2
\end{equation}

\noindent where $r_{jk}$ are the $\omega$-dependent complex Fresnel reflection coefficients of the different interfaces, the subscripts 1, 2 and 3 refer to vacuum, n-Ge film and Si substrate respectively, the subscript pair indicates the corresponding interface, $\eta(\omega)$ and $\alpha(\omega)$ are the n-Ge refractive index and absorption coefficient and $d_{\text{IR}}$ is the IR thickness which, at first order, equals the physical thickness. The backside substrate-vacuum interface is not included in Eq.\ \ref{model} because it is left unpolished. In the far-IR, for large enough value of the product $\alpha(\omega)d_{\text{IR}}$, one has $R(\omega) \simeq |r_{12}|^2$ and the Si substrate properties, which enter only in $r_{23}$, do not contribute to the reflected intensity spectrum. The condition $\alpha(\omega)d_{\text{IR}} \gg 1$ was self-consistently checked after the determination of the dielectric function (see inset of Fig\ \ref{fig:epsilon}-c). For $\omega $  above the plasma edge ($\sim 1000$ cm$^{-1}$), the condition $\alpha(\omega)d_{\text{IR}} \gg 1$ does not hold and the dielectric function of the Si substrate measured in a separate reflection/transmission experiment was used as a fixed input to reduce the number of free parameters. For $\omega > 1000 $ cm$^{-1}$, we found $\epsilon_{Si} \sim 11.9 + 0.1i$ for our wafers. The unknown dielectric function of the n-Ge film is modeled with a multi-oscillator Drude-Lorentz expression:

\begin{equation}
\label{lorentz}
\tilde{\epsilon}(\omega)=\epsilon_\infty-\frac{\omega_p^2}{\omega^2+i\omega\gamma_{\text{D}}} + \sum_{i=1}^2 \frac{S_i^2}{(\omega_i^2-\omega^2)-i\omega\gamma_i}
\end{equation}

\noindent from which all the response functions in Eq.\ \ref{model} ($r_{12}, r_{23}, \eta, \alpha$) are determined by classical electrodynamics relations \citep{Dressel}. A fit to Eq.\ \ref{lorentz} provides the unknown free carrier parameters (the \emph{Drude weight} $\omega_p^2$ and the \emph{Drude scattering rate} $\gamma_{\text{D}}$) while $\epsilon_\infty = 16.0$. The frequency of the interband transitions is approximately known ($\hbar\omega_1 =0.66$eV for the indirect gap transition, and $\hbar\omega_2 = 0.84$eV for the direct gap transition at room $T$), and the parameters $S_1$, $S_2$ and $\gamma_1$, $\gamma_2$ assume almost the same values for all samples. Notice that the \emph{unscreened} plasma frequency $\omega_p$ is now defined as the square root of the Drude weight, and it is conceptually different from $\omega^*$ defined previously. $\omega_p$ is the absorption cross section of the free carriers and it is directly proportional to $\sqrt{n/m^*}$. $\omega^*$ is the highest frequency where metallic behavior is observed ($\epsilon^{\prime}<0$) hence it also depends on the screening field produced by the valence electrons, which is approximately accounted for by using $\epsilon_\infty > 1$. 

As explained in the introduction, n-Ge features a band structure that allows for an approximate description of its dielectric response in terms of the Drude term only (the first term of Eq.\ \ref{lorentz}), as is usually undertaken in the applied physics literature \citep{Arizona}. Small deviations from the Drude-only model are expected only at $\omega$ well above the plasma edge, due to the low-energy tails of the weak interband transitions centered at $\omega \geq 5300 $ cm$^{-1}$. The consequence of simplifying the Drude-Lorentz model to an effective Drude-only model is the use of a phenomenological value for the infinity dielectric constant that we accurately determined to be $\epsilon_\infty^* = 16.6 > 16.0$ in our case. The slightly higher value of $\epsilon_\infty^*$ if compared to the nominal intrinisc-Ge squared refractive index value 16.0 accounts for the extra-screening of free-carrier oscillations by dipole moments of the interband transitions, which are slightly stronger in doped materials because crystal defects enhance the dipole moment of the indirect-gap transition \citep{amorphous}. Samples with high crystal defect densities and strongly enhanced interband transition strengths have demonstrated values of $\epsilon_\infty^*$ up to 18 (not reported here). Noticeably, the best-fit Drude-only parameters $\omega_p$ and $\gamma_{\text{D}}$ coincide within errors (around $\pm$ 2\%) with those of the Drude-Lorentz fit. A more accurate value of $\omega^*$ at room $T$ can then be obtained from the condition $\epsilon^{\prime}(\omega^*)=0$ that simply gives:

\begin{equation}
\label{screened2}
\omega^* = \sqrt{\frac{\omega_p^2}{\epsilon_\infty^*}-\gamma_{\text{D}}^2}
\end{equation}

\noindent and provides a direct estimate of $\omega^*$ from the Drude-only fitting parameters. Only in the zero loss limit $\omega_p \gg \gamma_{\text{D}}$, the above expression reduces to the approximate relation for the screened plasma frequency (Eq.\ \ref{screened}). One can see that, due to losses, the difference between Eq.\ \ref{screened2} and Eq.\ \ref{screened}, the latter commonly employed in the plasmonics literature \citep{BoltassevaRev}, is not negligible in many of the materials considered for mid-IR plasmonics applications.

\subsection{Infrared estimate of the free carrier density}
We now turn to the determination of the free carrier concentration $n$ from the Drude weigth $\omega_p^2$:

\begin{equation}
\label{unscreened}
n_{\text{IR}} = \frac{\omega_p^2 m^*}{4 \pi\epsilon_0 e^2}.
\end{equation}

The determination of the zero-crossing frequency of $\epsilon^{\prime}(\omega)$ by mid-IR spectroscopic ellipsometry and the subsequent retrieval of $\omega^2_p$ through Eq.\ \ref{screened2} by using estimates for $\epsilon_\infty^*$ and $\gamma_{\text{D}}$ \citep{Arizona}. We have already seen, however, that the estimate of $\epsilon_\infty^*$ is not straightforward even in the case of well-known semiconductors like Ge. We will also see below that, in the case of doped semiconductors, the parameter $\gamma_{\text{D}}$ represents just a rough estimate of the losses. Instead, fitting either the Drude-Lorentz or the Drude-only model to the entire far- to near-IR $R(\omega)$ almost eliminates the dependence of the relevant parameter $\omega_p$ on $\epsilon_\infty^*$ and $\gamma_{\text{D}}$.
The uncertainty of $n_{\text{IR}}$ is around $\pm$4\% propagated from the uncertainty in the fitting parameter $\omega_p$ of $\pm$2\%, and it is not limited by the smaller uncertainty on $m^*=0.12$ \citep{levinger1961}. The $n_{\text{IR}}$ values determined from the Drude-only fit are reported in Table \ref{tab:table1} and \ref{tab:table2}. One can see from Table \ref{tab:table1} that the value of $n_{H}$, which is determined at low $T$, well matches the value of $n_{\text{IR}}$ determined at room $T$ from the Drude weight. This indicates that both values are weakly $T$-dependent, as typically found for bulk crystals and epitaxial structures of high-purity semiconductor materials with doping level beyond the Mott transition. Our hetero-epitaxial n-Ge films are no exception. $n_{\text{IR}}$ is also less dependent on the substrate conductivity than $n_H$ and therefore it is used to determine the activation ratios reported in Table \ref{tab:table1}. A further model-independent method to determine $\omega_p$ and then $n$ from the IR data is provided by the oscillator strength sum rules \citep{Dressel}, which can be applied only once the optical constants are calculated in the entire IR range by the Kramers-Kronig (KK) transformations, as done in Ref.\ \onlinecite{Ortolani2005} and below in this work.

\subsection{Infrared estimate of the scattering time}
We now turn to the comparison of $\tau_H$ with the Drude scattering time $(2\pi c \gamma_{\text{D}})^{-1}$ determined from the fitting routine at room $T$ and reported in Table \ref{tab:table2}, where $c$ is the velocity of light in a vacuum. While the values of the two quantities fall approximately in the same sub-picosecond range for all samples, there is a strong dependence of $\tau_H$ on $n$ that has no counterpart in $(2\pi c\gamma_{\text{D}})^{-1}$, which is doping-independent. The reason is that $\tau_H$ is mainly determined by the electron momentum-relaxation process through electron scattering events with vanishing energy exchange, including elastic scattering with impurities, while $(2\pi c\gamma_{\text{D}})^{-1}$ is the inelastic scattering time of oscillating electron currents, taken as constant in a broad range of IR frequencies of the order of $\gamma_{\text{D}}$ itself. The main inelastic mechanism is electron-phonon scattering that does not depend on the electron density and this explains the doping-independence of $\gamma_{\text{D}}$. It should also be noted that the scattering times $\tau_H$ and $(2\pi c\gamma_{\text{D}})^{-1}$ are determined from processes that display very different dependence on the scattering angle \citep{Ridley}. However, one could still argue that electron momentum and energy relaxation processes are linked to each other and therefore there should exist a limit for $\omega \to 0$ where optical and transport scattering time approximately coincide. Indeed, this limit exists and it is represented by the empirical Hagen-Rubens relation between the normal-incidence reflectivity and the dc conductivity $\sigma_{\text{dc}}=1/\rho$, which holds for good conductors in the frequency range where the imaginary part of $\tilde{\sigma}(\omega)$ is negligible if compared to the real part \citep{Dressel}:

\begin{equation}
\label{hr}
R(\omega) = 1- A(\omega) \simeq 1- \sqrt{A_{\text{HR}}\omega} = 1 - \sqrt\frac{2\omega}{\pi\sigma_{\text{dc}}}  
\end{equation}

\noindent where $A(\omega) \simeq \sqrt{A_{\text{HR}}\omega}$ is the emittance of the semi-infinite medium \citep{Dressel}. We can fit Eq.\ \ref{hr} to our reflectivity data in the far-IR range $\omega \alt 200$ cm$^{-1}$ to determine the coefficient $A_{\text{HR}}$ (see inset of Fig.\ \ref{fig:drude}, from which we  calculate a low-frequency scattering time $\tau_{\text{IR}} = 8A_{\text{HR}} / \omega_p^2 $ using the value for the Drude weight derived from the Drude-Lorentz model, Eq.\ \ref{unscreened} and the relation $\sigma_{\text{dc}}=ne^2\tau/m^*$. As seen from Table \ref{tab:table2}, the value of $\tau_{\text{IR}}$ is similar to $(2\pi c\gamma_D)^{-1}$ but it decreases with increasing doping like $\tau_H$.

\subsection{IR estimate of the film thickness}
The best-fit value of $d_{IR}$ is reported in Table \ref{tab:table1} and it is mainly determined by the Fabry-Perot fringe pattern visible in $R(\omega)$ above the screened plasma frequency. The reason for the discrepancy of $\sim 15$\% between $d_{\text{IR}}$ and the physical thickness $d$ is unclear at the moment, and it could be due either to experimental uncertainties in $d$ or to non-ideality of the optical setup such as the finite distribution of incidence angles around the normal direction that affected the absolute level of $R(\omega)$, or both, as well as to the failure of the smooth optical interface model, or finally to the non-Lorentzian shape of the indirect-gap absorption feature at 5500 cm$^{-1}$.

\subsection{Low-temperature electrodynamics}
We have found that the Drude-Lorentz model reasonably reproduces the room-$T$ $R(\omega)$. The situation is completely different when one turns to the low-$T$ $R(\omega)$. If one looks at the data taken at $T = 300$ and $10$ K (thin colored lines in Fig.\ \ref{fig:drude}), one sees that the plasma edge does not shift appreciably, as expected from the fact that $n$ is constant with $T$. At the same time, the low-frequency value of $R(\omega)$ increases in the entire far-IR range with cooling, which is consistent with a reduction of the electromagnetic field penetration depth (skin depth) due to decrease of free carrier losses. The best-fit of Eq.\ \ref{model} to the $R(\omega)$ data at low $T$, however, based on the Drude model of Eq.\ \ref{lorentz}, is not possible, because there is no single value of the fit parameter $\gamma_{\text{D}}(10$ K) that reproduces both the increased far-IR reflectance (see also the inset of Fig.\ \ref{fig:drude}) and the substantial insensitivity to $T$ of the slope of the plasma edge in the mid-IR. This is clearly demonstrated by the dashed grey lines in Fig.\ \ref{fig:drude} obtained by imposing a value constant with $T$ for the product $\gamma_{\text{D}}\cdot\mu_H$: in a broad range of frequencies, indicated by the black arrows, there is no overlap between the data and the model. It is possible to reproduce $R(\omega)$ at low $T$ by decreasing the Drude weight and by adding several Lorentz oscillators in the far-IR representing photoionization of weakly bound charges. Decreasing the Drude weight at low $T$, however, would conflict with the $T$-independent free carrier density established from both the Mott criterion and the almost $T$-independent $n_H(T)$. 

The failure of the Drude model in reproducing the  $R(\omega)$ spectra at low $T$, together with the above mentioned inconsistency between $\tau_{\text{IR}}$ and $(2\pi c\gamma_{\text{D}})^{-1}$ at room $T$, indicate that the interpretation of energy relaxation using a single characteristic electron scattering time is not valid in heavily doped semiconductors, as already pointed out from theory \citep{SpitzerTheo}. Indeed, it is well known that the Drude model reproduces the IR data in limited IR frequency ranges only \citep{SpitzerExp}. Instead, for a correct electrodynamic description of electron scattering, one should use (at any $T$) the frequency-dependent scattering rate $\gamma(\omega)$. Conceptually, $\gamma(\omega)$ includes the dependence of the electron scattering cross section on energy and the frequency-dependent joint density of states (JDOS), i.e. the product of the density of occupied initial states and unoccupied final states separated by a given energy step. These effects cannot be ignored in heavily doped semiconductors, because their Fermi level is close to the minimum of a parabolic band leading to a strong dependence of the JDOS on the electron energy. Moreover, the electron-energy dependence of the scattering cross-section is strong in the case of (inelastic) optical phonon scattering. Therefore, the values of $\gamma_{\text{D}}$ used to reproduce a subset of the IR data within the Drude model are to be interpreted as the average value of $\gamma(\omega)$ in that specific frequency range (see e.\ g.\ Ref.\ \onlinecite{SpitzerExp}). In the far-IR, momentum and energy relaxation processes are degenerate and therefore one finds $\tau_{\text{IR}} < (2\pi c\gamma_{\text{D}})^{-1}$). In the next two subsections, we describe a procedure to determine $\gamma(\omega)$ at all frequencies from the IR data in a model-independent way.

\begin{figure*}
\includegraphics{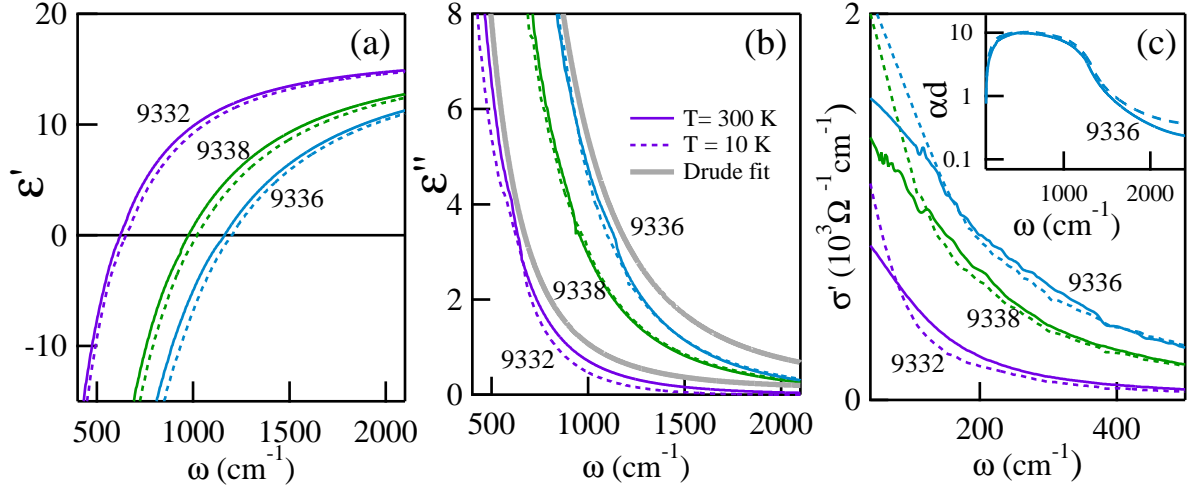}
\caption{\label{fig:epsilon} (Color online). (a,b) Dielectric function and (c) optical conductivity as determined by the Kramers-Kronig transformations for samples 9332, 9338 and 9336. Solid lines: $T=300$K, dashed lines: $T=10$ K. In panel (b), $\epsilon^{\prime\prime}(\omega)$ calculated with the Drude-lorentz formula and best-fit parameters for samples 9332 and 9336 is also shown (thick grey lines): clearly, the model does not reproduce the mid-IR data of sample 9332 (9336) for $\omega \agt 700 (1000)$ cm$^{-1}$. In the inset of panel (c), the product $\alpha(\omega)d_{\text{IR}}$ for sample 9338 is much larger than unity in the range of interest, validating the KK analysis based on semi-infinite medium approximation in a self-consistent way.}
\end{figure*}

\begin{table*}
\caption{\label{tab:table2}n-Ge parameters relevant for mid-IR plasmonics}
\begin{ruledtabular}
\begin{tabular}{ccccccccccc}
 \multicolumn{4}{c}{$T$-independent}&\multicolumn{3}{c}{$T$ = 300 K}&\multicolumn{3}{c}{$T$ = 10 K}\\
 Sample& $\omega_p$(cm$^{-1}$)& $n_{\text{IR}}$(cm$^{-3}$)&$(2\pi c\gamma_{\text{D}})^{-1}$(ps)&$\omega^*$(cm$^{-1}$)&$\tau_{\text{IR}}$(ps)&$\tau_{H}$(ps)&$\omega^*$(cm$^{-1}$)&$\tau_{\text{IR}}$(ps)&$\tau_H$(ps) \\ \hline
8648&$ 390$ & $2.1\cdot 10^{17}$&$0.026$&$-$& $-$&$0.064$&$-$ & $-$&$-$\\
8643&$1600$ & $3.5\cdot 10^{18}$&$0.026$&$310$&$0.025$&$0.032$&$-$ & $-$&$-$\\
8649&$2840$ & $1.1\cdot 10^{19}$&$0.026$&$666$&$0.015$&$0.020$&$-$ & $-$&$-$\\
8644&$3310$ & $1.5\cdot 10^{19}$&$0.026$&$815$& $0.014$&$0.012$&$-$ & $-$&$-$\\
9007&$4280$ & $2.5\cdot 10^{19}$&$0.017$&$1060$& $0.014$&$0.020$&$-$ & $-$&$-$\\
\hline
9332&$2450$ & $0.8\cdot 10^{19}$&$0.026$ &$620$ &$0.027$&$0.032$&$650$ & $0.055$&$0.046$\\
9338&$4200$ & $2.3\cdot 10^{19}$&$0.021$&$970$ &$0.019$&$0.020$&$1020$ & $0.032$&$0.032$\\
9335&$4230$ & $2.5\cdot 10^{19}$&$0.021$&$1050$ &$0.018$&$0.020$&$1070$ & $0.036$&$0.025$\\
9336&$4800$ & $3.0\cdot 10^{19}$&$0.021$&$1160$ &$0.018$&$0.016$&$1200$ & $0.024$&$0.018$\\
\end{tabular}
\end{ruledtabular}
\end{table*}

\subsection{Kramers-Kronig transformation analysis}
In the Kramers-Kronig (KK) transformations, the IR reflectance data are analyzed in a model-independent way. The film is assumed to be semi-infinite and the phase $\theta(\omega)$ of the Fresnel reflection coefficient $r_{12}(\omega)$ is retrieved by applying KK transformations to $|r_{12}(\omega)|=\sqrt{R(\omega)}$. With the complex semi-infinite medium reflectance $\tilde{r}_{12}(\omega)$ at hand, one can calculate all other optical response functions of n-Ge as a function of frequency \citep{Dressel}, including the optical conductivity $\tilde{\sigma}(\omega)$, the dielectric function $\tilde{\epsilon}(\omega)$ and the memory function $\tilde{\gamma}(\omega)$ (see below). To perform the KK analysis one needs to extend the $\omega$-dependence of the data outside the experimentally available range. For $\omega \to 0$, we extrapolated the merged $R(\omega)$ datasets with the Hagen-Rubens formula of Eq.\ \ref{hr}. For $\omega \to \infty$, we substituted the $R(\omega)$ data above the plasma edge (where the film becomes partly transparent) with the Drude-Lorentz model reflectivity calculated from Eq.\ \ref{model} in the limit $d_{\text{IR}} \to \infty$. This procedure is explained in details in Ref.\ \onlinecite{irmmw2014}. The resulting optical constants are accurate to within 5\% in the frequency range $(2\pi c\tau_{\text{IR}})^{-1} < \omega < \omega^*$. In Fig.\ \ref{fig:epsilon} we plot the resulting dielectric function of the 933x series at $T=10$ K and $T=300$ K. The data of sample 9335 are not reported for clarity as they overlap to those of sample 9338. Dashed lines are the room-$T$ values while continuous lines are the low $T$ values: it is apparent that, while the real part $\epsilon^\prime(\omega)$ shows a small $T$-dependence, the imaginary part $\epsilon^{\prime\prime}(\omega)$ at the two $T$'s almost overlap, indicating that the decrease of the mobility with cooling has a very weak effect on the mid-IR dielectric function close to $\omega^*$. This fact is better highlighted by plotting in Fig.\ \ref{fig:epsilon}-c the real part of the optical conductivity $\sigma^\prime = (-i\omega\epsilon_0/4\pi)\cdot \epsilon^{\prime\prime}$, which only at low $\omega$ is significantly $T$-dependent. The Drude weight $\omega_p$ determined from the Drude-only fit coincides within 5\% with the result of the oscillator strength sum-rule (frequency integral of $\sigma^\prime$ in Fig.\ \ref{fig:epsilon}-c) at both 300 K and 10 K. The data in Fig.\ \ref{fig:epsilon} differ considerably from the dielectric function obtained by the Drude-Lorentz and Drude-only model (see panel (b)). They can be considered an almost model-independent estimate of the dielectric function for n-Ge at all frequencies to be used for electromagnetic design. 

We can now summarize the optical parameters of n-Ge relevant for mid-IR plasmonics  in Table \ref{tab:table2}. The value of $\omega^*$ is here directly determined from the dielectric function resulting from KK analysis by searching the frequency where $\epsilon^{\prime}(\omega^*)=0$ in Fig.\ \ref{fig:epsilon}-a, instead of using the approximate relation of Eq.\ \ref{screened} or the model-dependent relation of Eq.\ \ref{screened2}. The values of $\omega^*$ at both 10 K and 300 K coincide within $\pm$5\% with the respective $T$-independent quantity $\omega_p/\sqrt{\epsilon_\infty}$, confirming that the Drude weight parameter of the Drude-Lorentz model provides a good estimate of the free carrier density, even if the model fails to accurately reproduce the dielectric function. Also, $\omega^*$ obtained from the KK analysis increases slightly with cooling, an effect that can be derived neither from Eq.\ \ref{screened}, where all quantities are constant with $T$ nor from Eq.\ \ref{screened2}, where $\gamma_{\text{D}}$ is undefined at low $T$. The weak but clear dependence of $\omega^*$ on $T$ reflects the slight decrease of electron energy losses in the mid-IR with cooling, due to the decrease of thermal population of optical phonons. Turning to the loss parameters, the Hagen-Rubens scattering time $\tau_{\text{IR}}$, determined from the zero-frequency extrapolation of $R(\omega)$, is approximately similar to $\tau_H$ and therefore is determined by both momentum energy relaxation processes with small energy exchange and energy relaxation processes. Instead, the Drude scattering time $(2\pi c\gamma_{\text{D}})^{-1}$ is longer than both $\tau_{\text{IR}}$ and $\tau_{H}$ because it is mainly determined by inelastic electron-phonon scattering (red arrow in Fig.\ 1), and not much by charged impurity scattering (indeed, $\gamma_{\text{D}}$ in Table \ref{tab:table2} is almost doping-independent). This fact suggests that $\gamma_{\text{D}}$ is not a good parameter to estimate plasmon losses which are certainly due also to momentum relaxation by charged impurity scattering (black arrow in Fig.\ 1). The estimate of mid-IR plasmon losses instead requires both the evaluation of the frequency-dependent scattering rate and the calculation of the total (elastic and inelastic) frequency-dependent average electron scattering cross sections, presented in the next sections.

\subsection{Frequency-dependent scattering rate}
The Drude model of the free carrier electrodynamics can be naturally extended to include the dependence of electron energy relaxation processes on $\omega$ by making the scattering rate of the Drude formula $\omega$-dependent (and therefore complex in order to satisfy its own KK relations). $\tilde{\gamma}(\omega)$ is one of the equivalent forms of the electrodynamic response function called the memory function. It can be demonstrated \citep{Allen1976, Goetze1981} that $\tilde{\gamma}(\omega)$ is related to $\tilde{\sigma}(\omega)$ as follows:

\begin{eqnarray}
\label{array}
\tilde{\gamma}(\omega)= \gamma^{\prime}(\omega) + i \gamma^{\prime\prime}(\omega) \nonumber \\
\gamma^{\prime}(\omega) = \frac{\omega^2_p}{4 \pi} Re \left(\frac{1}{\tilde{\sigma}(\omega)}\right) \nonumber  \\ 
\gamma^{\prime\prime}(\omega) = - \frac{\omega^2_p}{4 \pi\omega} \left[Im\left(\frac{1}{\tilde{\sigma}(\omega)}\right)-1\right] \nonumber \\  
\end{eqnarray}

\begin{figure*}
\includegraphics{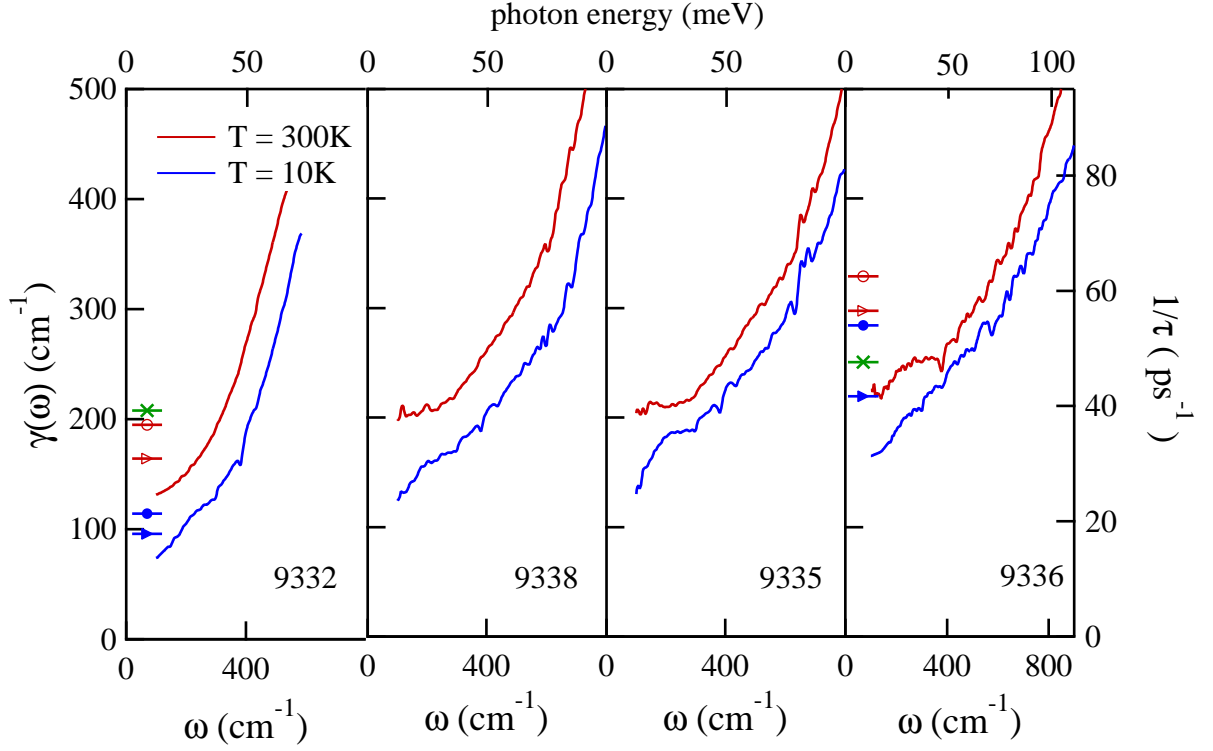}
\caption{\label{fig:Gamma}(Color online). Experimental frequency-dependent electron scattering rate. the symbols correspond to the values in Table \ref{tab:table2} for $\gamma_{\text{D}}$ (green cross), $\tau_{\text{IR}}$ (triangles) and $\tau_H$ (circles). Empty symbols are for room-$T$ values.}
\end{figure*}

This so-called extended Drude model has been employed to analyze the electrodynamics in the IR range of metals \citep{Allen1971, Shulga1991, vanderEb2001, Basov2002}, transition-metal compounds \citep{Allen1976}, heavy-fermion systems \citep{Degiorgi1999}, and high critical temperature cuprate superconductors \citep{Puchkov1996}. The real part of the memory function $\gamma^\prime(\omega)$ represents the $\omega$-dependent electron scattering rate averaged over all electrons occupying states within an energy shell $\pm\hbar\omega$ from the Fermi level, and normalized by the electron density. As such, $\gamma^\prime(\omega)$ can be used for an approximate evaluation of the energy relaxation rate of a finite-frequency collective electron excitation (plasmon at $\omega$) interacting with the crystal lattice. Writing $\tilde{\sigma}(\omega) = (4\pi i/\omega)^{-1} (\tilde{\epsilon}(\omega) -\epsilon_\infty)$, one can express $\gamma^\prime(\omega)$ in terms of the dielectric function:

\begin{equation}
\label{gamma}
\gamma^\prime(\omega)  
= \frac{\omega^2_p}{\omega} \frac{\epsilon^{\prime\prime}(\omega)}{(\epsilon^{\prime}(\omega)-\epsilon_\infty)^2+(\epsilon^{\prime\prime}(\omega))^2}
\end{equation}

\noindent where we can input the data shown in Fig.\ \ref{fig:epsilon} and obtain an experimental $\gamma^{\prime}(\omega)$ to be compared with the $\omega$-dependent scattering cross-section determined from first principles (see next section). For completeness, we mention that the imaginary part of $\tilde{\gamma}(\omega)$ represents the frequency-dependent effective-mass renormalization factor. The relation between $\gamma^{\prime\prime}(\omega)$ and $\tilde{\epsilon}(\omega)$ can be derived similarly to Eq.\ \ref{fig:Gamma} but it is not of interest here.

In Fig.\ \ref{fig:Gamma}, we plot $\gamma^\prime(\omega)$ at $\omega \alt \omega^*$ for the four samples that were studied as a function of $T$. It increases with frequency at both room $T$ and low $T$ for all doping levels, due to both the higher JDOS for processes with high energy exchange $\pm\hbar\omega$ and the increase of the electron-phonon scattering cross section with increasing electron energy. The increase of $\gamma^\prime(\omega)$ with $\omega$ is very important for the evaluation of mid-IR plasmonic losses, because it indicates that they are generally under-estimated when the scattering times determined from dc transport are employed. In Fig.\ \ref{fig:Gamma}, for samples 9332 and 9336, the values of $\tau_H^{-1}$ and $\tau_{\text{IR}}^{-1}$ (symbols) are lower by a factor of 2 or more if compared to the mid-IR scattering rates for $\omega>500$ cm$^{-1}$. The actual value of the electron scattering rates to be used for evaluation of plasmon losses is even higher than $\gamma^\prime(\omega)$, because the latter quantity is mainly determined by inelastic scattering events, while a plasmon can decay also through elastic scattering events that do not conserve the direction of the electron momentum (black arrow in Fig.\ \ref{fig:plasmon}). The estimate of elastic scattering rates cannot be done by IR spectroscopy and here it is done by first-principle calculations reported in the next section. We point out two further experimental facts: ($i$) $\gamma^\prime(\omega)$ is always higher at room $T$ than at low $T$ for each sample, indicating the key role played by absorption of thermally excited phonons; ($ii$) $\gamma^\prime(\omega)$ does not decrease in sample 9336 which is homoepitaxially grown on a Ge wafer hence has fewer crystal defects if compared to the other samples grown on Si wafers. Therefore it does not depend on the crystal defect density. These experimental facts help simplifying the first-principle calculations because they can be performed by taking into account only the interaction of electrons with optical phonons and charged impurities.

\section{Model of electron scattering}

The full evaluation of the electron scattering rate including elastic and inelastic processes, and without any dependence on the properties of the specific experimental probe, can be obtained from first-principles calculations of electron interactions with the crystal field or the impurity field in a three-dimensional solid-state environment. We consider a gas of free Landau quasi-particles with renormalized electron-electron interactions and conductivity effective mass $m^*=0.12 m_e$. For the band structure of germanium, we use an anisotropic parabolic-band approximation for the $L$-valleys, with two relevant values for the density-of-states effective mass $m_L=1.59 m_e$ and $m_T=0.0815 m_e$. Conduction-band minima other than the $L$-valleys become important only for $\omega>1100$ cm$^{-1}$, which corresponds to the energy separation between $L-$ and $\Gamma-$valley minima of 0.14 eV, and they are neglected in this work. The possible scattering channels for a single electron are due to the interaction with phonons (acoustic and optical branches), with charged impurities (activated dopant atoms of density $N_A$), with neutral impurities (inactivated dopant atoms of density $N_I$) and with crystal defects. We now analyze the different interactions and motivate our choice of including or not including them in our calculation.

\textbf{Charged impurities.} Coulomb scattering by charged impurities is known to be the main cause of reduction of  dc electron mobility in heavily-doped semiconductors, because the density of ionized donors acting as scattering centers is very high if compared with the intrinsic electron mean free paths observed in lightly-doped or remotely-doped materials, which are much longer than those found in metals or oxides.  We calculate the contribution of charged impurity scattering following Ref.\ \onlinecite{Ridley} and using the experimental value of $N_A$ for each sample.  At a first look, one may think that increasing the doping level may produce an increase in the scattering rate because $N_A$ increases. $n$ also increases, however, thereby reducing the screening length for each impurity. Interestingly, in the present range of doping of n-Ge around $2\cdot10^{19}$ cm$^{-3}$, this balance results in an almost doping-independent magnitude of the charged impurity effect, although the specific functional form of the scattering rate on the electron energy is different for each doping level. At first order, charged impurity scattering does not depend on $T$, as long as $N_A$ does not vary with $T$. The screening field has a space-time structure, however, therefore the single-electron cross section $p_{\text{cha}}(\omega)$ depends on the electron energy separation from the Fermi level $\hbar\omega$ only through the modulus $q$ of the momentum exchanged in the interaction, which is conserved during the scattering event \citep{Ridley}: 

\begin{equation}
\label{charged}
p_{\text{cha}}(q) \sim \frac{2\pi e^4 N_A }{\hbar\epsilon_\infty^2(q^2+q_{TF}^2)} 
\end{equation}
 
\noindent where $q_{TF}$ is the Thomas-Fermi inverse screening length and the $\omega$-dependence of $p_{\text{cha}}(\omega)$ is calculated for each $q$ imposing energy conservation in the transitions in the anisotropic conduction band of Ge. We recall that in an isotropic parabolic minimum one could write the simple expression $q^2=4m^*m_e\hbar\omega(1-\cos\theta)$ where $\theta$ is the scattering angle, but not in n-Ge.

\textbf{Optical phonons.} The non-polar lattice of Ge does not allow for the existence of IR-active optical phonon modes whose dipolar field directly interacts with free carriers \citep{Ridley}. Indeed, $\epsilon^{\prime\prime}(\omega)$ in Fig.\ \ref{fig:epsilon} does not presents any absorption resonance at the optical phonon frequencies. In an ideal non-polar crystal, however, there is a residual electromagnetic interaction at the microscopic level between the free carriers and the atoms when they are located outside their equilibrium position e.g. by thermal fluctuations (deformation potential scattering \citep{Ridley}). At low $T$ where thermal fluctuations are small, the deformation potential scattering becomes relevant only if the electron has energy high enough to excite an optical phonon. The phonon should connect two electronic states by conserving total energy and momentum. Calling $\bf{k}$ the initial crystal momentum, $\bf{q}$ the phonon momentum, $\Delta V_{\textbf{q}}$ the variation of the band structure potential due to a perturbation of the crystal lattice with periodicity $2\pi/\bf{q}$ and $\varepsilon_{\bf{k}}$ the energy of electronic states, the single-electron scattering cross-sections for phonon absorption and emission can be written as:

\begin{eqnarray}
\label{phonon}
p^+_{\text{pho}}(\textbf{q},\omega) = \frac{<\textbf{k+q}|\Delta V_{\textbf{q}}|\textbf{k}>}{\varepsilon_{\textbf{k+q}}-\varepsilon_{\textbf{k}}} \\ \nonumber
p^-_{\text{pho}}(\textbf{q},\omega) = \frac{<\textbf{k-q}|\Delta V_{\textbf{q}}|\textbf{k}>}{\varepsilon_{\textbf{k}}-\varepsilon_{\textbf{k}-\textbf{q}}} \\ \nonumber
\end{eqnarray}

\noindent with the condition $\hbar\omega=\varepsilon_{\bf{k+q}}-\varepsilon_{\bf{k}}$ $(=\varepsilon_{\bf{k}}-\varepsilon_{\bf{k-q}} $) and with the constraint that $\omega(\bf{q})$ must correspond to an optical phonon branch of the material. In Ge, there is only one optical phonon mode with vanishing wavevector $\textbf{q}$ relevant for intravalley scattering of free carriers with small energy exchange $\hbar\omega$, i.\ e.\ the zone center mode with symmetry $\Gamma_{25}$, which is also Raman-active. Due to the cubic symmetry of the Ge lattice, transverse (TO) and longitudinal (LO) optical modes are degenerate at the $\Gamma$ point. The energy of this mode is $\omega_{ph1}=297$ cm$^{-1}$ or $E_{ph1} = \hbar\omega_{ph1}=$37 meV. At odds with $n$-doped III-V compound semiconductor materials that have a single-valley Fermi surface at the $\Gamma$ point, Ge has 4 valleys at the $L$-points, therefore there is another relevant phonon mode for electron scattering with wavevector matching the distance in $\bf{k}$-space between the $L$-valleys. This inter-valley mode has $\omega_{ph2}=201$ cm$^{-1}$ or $E_{ph2} = \hbar\omega_{ph2}=$25 meV. The single-electron scattering cross-section for phonon absorption with phonons has a strong dependence on $T$, because it depends on the phonon population at a given $T$ according to the Bose-Einstein distribution function. Instead, at first order, electron-phonon scattering does not depend on the doping level.

\begin{figure*}
\includegraphics{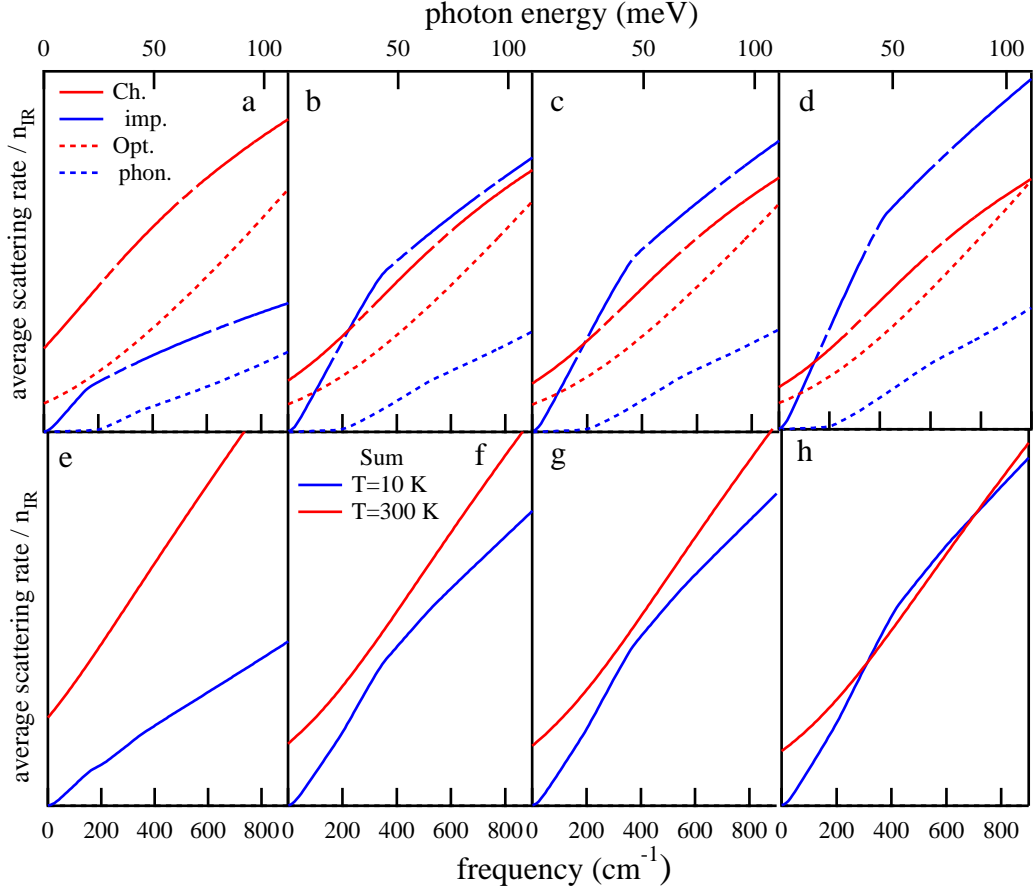}
\caption{\label{fig:theo}(Color online). The calculated frequency-dependent scattering rate averaged over all electrons in the $L$-minima of n-Ge with Fermi level giving an electron density equal to $n_{\text{IR}}$. Single-particle cross sections are averaged over all possible transitions, considering the effect of electron state occupation at all $T$ with the Fermi-Dirac function. All curves are normalized by the respective $n_{\text{IR}}$. (a-d) The contributions from charged impurity scattering (dash-dotted lines) and optical phonon scattering (dotted lines) are plotted separately. (e-h) The corresponding sum of the two scattering contributions to be compared with the experimental data in Fig.\ \ref{fig:Gamma}.}
\end{figure*}

\textbf{Acoustic phonons}. Interaction of electrons with acoustic phonons is a minor effect when the energy of the electron above the Fermi level is high enough to emit an optical phonon, i.e. for $\omega \agt 200$ cm$^{-1}$. Therefore, we expect that our calculations will provide accurate values for mid-IR scattering rates even if we neglect acoustic phonon scattering, which would be rather computation-intensive due to the large number of branches and of possible events. Note that an estimate of the effect of acoustic phonon scattering could be derived from a detailed analysis of the data in Fig.\ \ref{fig:hall} \citep{Mirza2014}, and also by looking at the experimental curves of Fig.\ \ref{fig:Gamma}. Therein, the deviation of $\gamma^\prime(\omega)$  below 200 cm$^{-1}$ from the almost-linear dependence on $\omega$ seen in the mid-IR, more visible at room $T$, can be taken as a signature of acoustic phonon scattering. Indeed, this is not seen in the calculations of Fig.\ \ref{fig:theo} that consider only the effect of optical phonons.

\textbf{Neutral impurities and defects}. Neutral impurities scatter electrons through static deformation of the periodic potential that forms the electronic band structure. Inactivated dopants are certainly present in our samples with high density $N_I$ of the same order of magnitude of $N_A$, but they contribute to electron scattering to a much smaller extent than charged impurities because their interaction with electrons is far weaker \citep{Mirza2014}. Neutral crystal defects such as stacking faults also contribute to electron scattering, but their average distance is, in the worst-case of sample 9338, in the range $\sim 0.1\rightarrow 1.0 \mu$m, much longer than the electron mean free path $\ell_{3D}$ of few tens of nanometers, obtained from $\ell_{3D} = (\hbar\mu_H/e)(3 \pi^2 N_A/2)^(1/3)$ \citep{Mirza2014} (we find $\ell_{3D} = 19$ nm for sample 9338 at 300 K). We can therefore neglect the effect of neutral impurities in the calculation.

\textbf{Total scattering rate}. We evaluated the single-particle scattering rates for charged impurity scattering using parameters for electrons at the $L$-point of the conduction band of Ge for four different values of charged impurity density $N_{A}$, which was taken equal to $n_{\text{IR}}$ to reproduce the situation of each of the four samples of the 933x series. $N_{A}= n_{IR}$ also defines the value of the Fermi level $E_F$ for each sample in the parabolic-band approximation ($E_F$ ranges from 20 to 40 meV for the 933x series). The total frequency-dependent scattering rate averaged over all electrons $\gamma_{\text{theo}}(\omega)$ is then obtained for each sample by summing the single-particle scattering cross-sections over all pairs of initial occupied states $|o>$ and final empty states $<e|$:

\begin{equation}
\label{fermi}
\gamma_{\text{theo}}(\omega) = \sum_{<e|,|o>} p_{\text{cha}}(\omega) + \sum_{<e|,|o>} p_{\text{pho}}(\omega)
\end{equation}

\noindent where the first and the second term represent elastic and inelastic scattering, respectively. The occupation probability of the states at each $T$ and for each value of the Fermi level (set by $N_A$) is calculated by the corresponding Fermi-Dirac distribution functions. Thus, the above expression for the total average scattering rate introduces in $\gamma_{\text{theo}}(\omega)$ further dependence on both doping and $T$ with respect to the single-particle scattering cross-sections $ p_{\text{ch}}(\omega)$, which is $T$-independent, and $p_{\text{el-ph}}(\omega)$, which is doping-independent. The overall tendency of the experimental scattering rate curves plotted in Fig.\ \ref{fig:Gamma} is fully captured by the calculation: both theoretical and experimental scattering rates increase with frequency, are higher at room $T$ than at low $T$ in all samples, and slightly increase with doping at all $T$'s. The zero-frequency value of $\gamma_{\text{theo}}(\omega)$ at 300 K is due to the optical phonon population only, while that in the experimental data of Fig.\ \ref{fig:Gamma} is due to both acoustic and optical phonons. The different functional form of $\gamma^{\prime}(\omega)$ in Fig.\ \ref{fig:Gamma} and of $\gamma_{\text{theo}}(\omega)$ in Fig.\ \ref{fig:theo} is to be ascribed to the contribution of elastic scattering by charged impurities, which displays a sub-linear dependence on $\omega$. The absolute values of $\gamma_{\text{theo}}(\omega)$ are calibrated with the experimental value for sample 9338 at $T=300$ K and $\omega=800$ cm$^{-1}$ where electron-phonon scattering is likely to be dominant.

\begin{figure}
\includegraphics{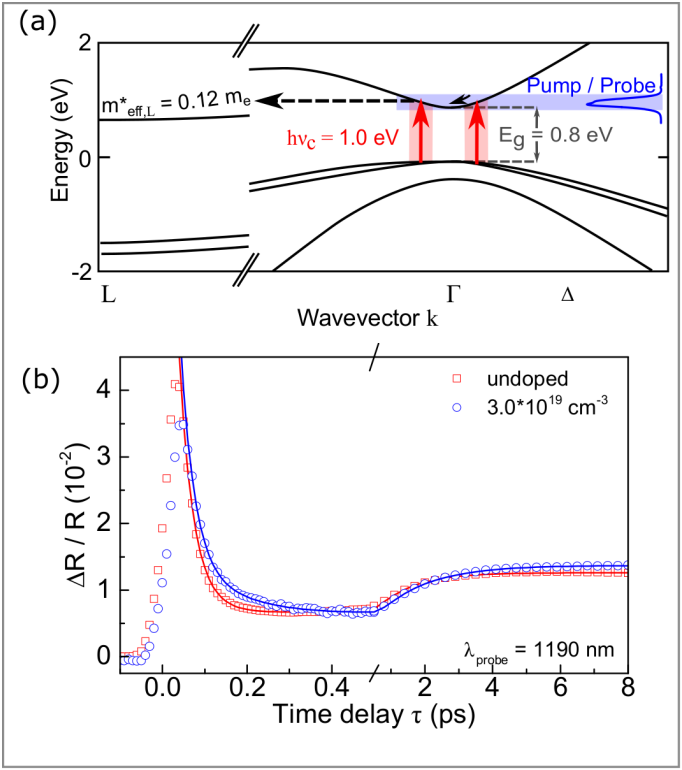}
\caption{\label{fig:pump}(Color online). Figure 10 – a) A simplified band structure of Ge depicting the impulsive excitation of electrons with a 20-fs pump pulse. $h\nu_c$ is the excitation photon energy and $E_g$ the direct bandgap. The arrows represent the scattering processes 1 and 2 occurring on a sub-ps timescale (see text): electron thermalization in $\Gamma$ (solid arrow) and intervalley scattering (dashed arrow). b) The transient reflectivity in the two samples as a function of time delay. Squares and circles depict the experimental data for the undoped and doped specimens respectively. Solid lines are the results of the fitting with a tri-exponential function. Notice the change of scale in the horizontal axis.}
\end{figure}

\section{Time-resolved spectroscopy}

Time-resolved transient reflectivity measurements were conducted on two Ge films grown on silicon substrates: one nominally identical to 9007 and one undoped. The experimental setup employs an optical parametric amplifier driven by an Yb:KGW regenerative laser system. This source delivers 20 fs near-IR pulses with 50 kHz repetition rate and wavelength components between 1000 and 1400 nm. The experiments are conducted in a degenerate geometry and the beam is split in two replicas, one used for resonant excitation of electrons from valence to conduction band at the $\Gamma$ point and one to probe the normal reflectivity change at a variable delay \citep{femtopump}. The pump spectrum allows for the selective excitation of direct transition in Ge while leaving unperturbed the Si substrate \citep{steadypump}. The transient reflectivity is acquired by spectrally filtering the probe pulse at the wavelength 1190 nm with a monochromator equipped with an InGaAs photodiode. The detection is based on a lock-in scheme with mechanical modulation of the pump beam with a chopper wheel.
The excitation fluence is kept at the moderate value of 500 $\mu$J/cm$^2$ to minimize two-photon absorption that would directly create electron-hole pairs at the $L$-point \citep{NJPhys}.

Figure \ref{fig:pump}-a sketches the pump-probe experiments in Ge. The electronic distribution that follows the 20-fs pump pulse is centered at the direct gap where the electrons display excess energy thus being in a nonequilibrium state. The evolution of this perturbed system is determined by four main physical processes \citep{thermalization}: (i) thermalization in the $\Gamma$ valley mainly via electron-electron scattering processes, this process is extremely fast and leads the electronic subsystem to a well-defined hot temperature; (ii) intervalley scattering of electrons to populate the $L$-valley; (iii) cooling of the electrons in the $L$-valley to reach an equilibrium with the lattice temperature; (iv) electron-hole recombination. The latter process occurs on a relatively long timescale in an indirect semiconductor as Ge and as such is not addressed by this study. The first three mechanisms are instead extremely interesting since they are related to the same fundamental interactions between electrons and lattice and between electrons and impurities that determine the plasmonic losses.

Figure \ref{fig:pump}-b shows the experimental traces of transient reflectivity as a function of time delay between pump and probe. The largest contribution to the signal occurs at early times for both samples. In fact, upon photoexcitation the semiconductor bleaches via Pauli blocking until the electrons are in the center of the Brillouin zone. Subsequently, the signal evolves via three time scales assigned to the first three processes discussed above: $\tau_1$ is associated with the thermalization process within the $\Gamma$ valley, $\tau_2$ with the intervalley scattering and $\tau_3$ with the electron cooling in the $L$-valley. Of course one has $\tau_1<\tau_2<\tau_3$. Their value is retrieved by fitting a tri-exponential function to the experimental data. The values obtained for the two fast processes are nearly independent on doping ($\tau_1=33$ fs for both samples and $\tau_2=194 (206)$ fs for the undoped (doped) sample). It should be noted that these experimental values are due to the reflectivity change at one specific probe photon energy. Their value is only related to the characteristic scattering time of one specific class of processes. In fact, electron scattering contributes to the reflectivity change only after a number of events large enough to significantly change the electron distribution.
Interestingly, the characteristic time $\tau_3$, which is related to cooling within the $L$-valley, differs significantly between the two samples. It is longer for the undoped sample ($\tau_3=1.4$ ps) than in the doped one ($\tau_3$ =1.0 ps), because additional scattering channels are activated by the presence of charged impurities. In fact, the combination of the two early processes results in a hot-electron population in the $L$-valleys having an energy spread of few hundreds of meV. This hot-electron population relaxes to quasi-equilibrium with the lattice by two distinct scattering mechanisms: one involving the phonons, and the other one involving the charged impurities and the (cold) electrons already present in the $L$-valleys due to static doping. 

The energy relaxation mediated via scattering with phonons is of high interest here, because it shares some features with the energy relaxation time of a collective plasmon excitation of energy $E_{pl}= k_BT_{e,0}$ where $T_{e,0} > 300$ K is the electron temperature in the $L$-valley at time delay equal to $\tau_2$. We can therefore define a plasmon energy relaxation time $\tau_{pl}$ and tentatively identify it with $\tau_3$ of the undoped sample. The energy lost by one electron in one single inelastic scattering event may be smaller than its total excess energy, and events resulting in electron energy increase (phonon absorption) can also occur. Therefore, many single-electron scattering events are needed for cooling the entire electron population and reach the equilibrium with the lattice, and we can expect $\tau_{pl} \gg (2\pi c \gamma(\omega))^{-1} $ as  actually found ($\tau_3 \sim 1.4$ ps $\gg (2\pi c \gamma(\omega))^{-1} \alt 0.05$ ps suggesting that a number of single electron scattering events $\agt 30$ is needed for plasmon energy relaxation). In heavily doped samples, calculations shown in Fig.\ \ref{fig:theo}a-d indicate that the average electron scattering rate due to optical phonons alone and to the presence of charged impurities should be comparable. The increased energy relaxation efficiency would be consistent with a lifetime of plasmons in doped samples shorter by a factor of $\sim$ 2 if compared to what expected from electron-phonon scattering only. 

The above speculation is useful to get rough estimates of mid-IR plasmon lifetimes, but it does not allow us to conclude that $\tau_3$ is a direct measure of this quantity. In fact, the evolution in zero external field of the nonequilibrium electron distribution prepared by the pump pulse is not the same process as the damping of plasmons excited by a continuous-wave radiation source. Although the energy spread of the electron distributions may be comparable in the two cases, they are certainly different in terms of the momentum distribution (see sketch in Fig. 1). Moreover, while in steady-state plasmon excitation $N_A = n$, in the pump-probe experiment $N_A < n$ and a hole gas with density $p = n$ is also present. For these reasons, the average electron scattering rates are expected to be higher in the pump-probe case, therefore $\tau_3$ may represent only a lower bound for the actual plasmon decay time. Nevertheless, it can be stated that the pump-probe data are consistent with the steady-state spectroscopy data, in the sense that there is no contradiction in assuming that the same single-electron scattering mechanisms govern plasmon decay and hot electron relaxation after optical pumping. The n-Ge energy relaxation in the mid-IR range seems to be a much slower process than single-electron inelastic scattering and therefore at odds with group III-V semiconductors, because of the lack of polar optical phonons, as already observed in n-Ge quantum-well systems \citep{Ortolani2011}.

\section{Discussion}

We have observed that, due to the peculiar frequency dependence of the electron-phonon and electron-charged impurity scattering cross sections in doped semiconductors with Fermi level close to a parabolic band-structure minimum, the increase of the electron mean free path with cooling usually observed in experiments conducted from dc up to terahertz frequencies has almost no counterpart at mid-IR frequencies. This means that, with cooling, there is not much to be gained in terms of mid-IR plasmon losses in semiconductors. Also, the frequency dependence of $\gamma(\omega)$ described in this work explains why semiconductors displaying high dc mobility like Ge or InAs, when employed for plasmonics in the mid-IR, did not provide significant advantages in terms of  quality factor of plasmonic resonances \citep{law2013, baldassarre2015}. The positive result of the present study is that, by increasing the doping level beyond $n \sim 2\cdot10^{19}$ cm$^{-3}$ as required for mid-IR plasmonics, the plasmon decay time is not expected to decrease further, because of the increased efficiency of free-carriers in screening the charged-impurity scattering field for electrons oscillating at mid-IR frequencies. Once again, this behavior is in contrast with the clear drop in the dc mobility $\mu_H$ with increasing doping. Therefore, the dc mobility cannot be used as a relevant parameter for losses in mid-IR plasmonics.

The experimental data presented in this paper also allows one to draw some conclusion on the relation between the average electron scattering time and the surface plasmon decay time. In the hydrodynamic regime, the electron-electron collisions conserve the total momentum and energy of the electron system. The hydrodynamic hypothesis corresponds to the assumption that all electrons at a given position contribute to the surface plasmon excitation. In this condition, the electron-electron collisions do not change the total electron momentum and energy, therefore they do not affect the plasmon decay time. On the other hand, electron-phonon and electron-charged impurity scattering events do not conserve the total momentum and/or energy and therefore they are certainly responsible for the plasmon decay. We have found, however, that a large number of single-electron scattering events is required for plasmon decay in n-Ge, as suggested by time-resolved data, and decay times of few picoseconds may be expected for mid-IR plasmons in nanostructures. The shorter $\tau_3 \sim 1.0$ ps found for the doped sample if compared to $\tau_3 \sim 1.4$ ps for the undoped one is probably due to the additional charged impurity (elastic) scattering contribution that increases the efficiency of energy relaxation. A similar energy relaxation efficiency increase also leads to the doping-dependent increase of $\gamma(\omega)$ seen in Fig.\ \ref{fig:Gamma}. With a radiation period of 30 fs at $\lambda=10$ $\mu$m, one can therefore assume that, even in heavily doped samples at room $T$, several cycles of plasma oscillation will take place before complete energy damping.

Note that the band-structure considerations made in the introduction are based on the corpus of experimental and theoretical knowledge acquired in many decades on the physical properties of bulk Ge crystals and do not necessarily hold for Ge thin films produced with all growth methods, substrate types and thicknesses. The documented high-crystal quality of the Ge films grown by CVD studied in this work explains why, as far as mid-IR optical properties are concerned, the crystal defects introduced during epitaxial growth of Ge-on-Si are not relevant and the mid-IR response is identical to that reported in the literature for bulk Ge crystals at comparable doping level \citep{SpitzerExp}.

One effect that is not discussed in the present work is the impact of the  $\bf{k}$-space direction of electron scattering on the plasmon decay time. This topic is somehow related to the difference between the isotropic momentum relaxation of optical scattering events and the strong dependence of the dc electron mobility on the direction of the exchanged momentum (or scattering angle) in the scattering event. For example, inelastic scattering events that produce exchange of momentum in the direction transverse to the plasmon wavevector may not impact on the plasmon decay. Another effect to be addressed is the matching of both frequency and mode symmetry between plasmon and phonon oscillations (polaritonic effect), which may lead to faster resonant plasmon decay at specific frequencies and wavevectors. This resonance effect is well known for polar phonons in III-V semiconductors \citep{dekorsy} and it has been recently studied in the case of non-polar phonons of graphene with striking results \citep{tonylow}. Finally, acoustic phonon scattering has not been considered in our first-principle calculations and the density of states has been assumed to be perfectly parabolic.

Given the comparable contributions of electron-phonon and electron-charged impurity scattering to plasmon decay, one could speculate that the electron-hole plasma in optically pumped intrinsic germanium \citep{NJPhys} would display highly tunable $n$ together with plasmon decay times dominated by electron-phonon scattering hence twice longer than those of plasmons in doped materials with comparable static electron (or hole) density. Inter-valence band transitions, however, will be activated in an electron-hole plasma created by optical pumping: further studies are required in order to verify whether mid-IR plasmons in optically-pumped intrinsic semiconductors display higher, comparable or lower losses than those of plasmons in static heavily-doped materials.

\section{Conclusions}
In conclusion, we have investigated the electron scattering mechanisms in heavily n-doped germanium thin films grown on silicon wafers and intended for mid-IR plasmonics and metamaterial applications. Hall transport measurements and IR spectroscopy data interpreted within the Drude-Lorentz model are in good agreement with the estimate of the free carrier density, that can reach 3 $\cdot$ 10$^{19}$ cm$^{-3}$, which corresponds to a plasma frequency of $\sim 1200$ cm$^{-1}$ or a plasma wavelength of 8 $\mu$m. The plasmon decay times cannot be estimated by the dc mobility, nor within the Drude-Lorentz model, because we demonstrated inconsistency between the single-valued Drude scattering rate and the actual frequency-dependent scattering rate. We determined the latter quantity both experimentally, by Kramers-Kroning transformations, and theoretically, by first-principle calculations. We have found a comparable role of electron-phonon and electron-charged impurity scattering in n-Ge. The average electron scattering rate, in the range of 10 to 30 fs for heavily doped films, increases strongly with excitation frequency, weakly with temperature, and it is almost constant with doping. Interband pump-probe experiments on heavily-doped n-Ge films suggest that the decay time of collective (plasmon) excitation may reach few ps, allowing for underdamped mid-IR plasma oscillations even at room temperature.

\section{Acknowledgements}
The research leading to these results has received funding from the European Union’s Seventh Framework Programme under grant agreement no. 613055

\end{document}